\documentclass[a4paper]{aa}
\usepackage{epsfig}
\def\ergsec{\hbox{erg s$^{-1}$}}

\def\degmark{^\circ}
\def \rsun {\ifmmode$R$_{\odot}\else R$_{\odot}$\fi}
\def \nh {N${\rm _H}$}
\def \hcm {\hbox {\ifmmode $ atoms cm$^{-2}\else atoms cm$^{-2}$\fi}}
\def \src {4U\,1630--47}

\def\approxgt{\mathrel{\hbox{\rlap{\lower.55ex \hbox {$\sim$}}
        \kern-.3em \raise.4ex \hbox{$>$}}}}
\def\approxlt{\mathrel{\hbox{\rlap{\lower.55ex \hbox {$\sim$}}
        \kern-.3em \raise.4ex \hbox{$<$}}}}

\newcommand {\sax} {{\it BeppoSAX}}

\def\arcmin{\hbox{$^\prime$}}

\newcommand {\chisq} {$\chi ^{2}$}
\newcommand {\rchisq} {$\chi_{\nu} ^{2}$}

\newcommand{\mc}{\multicolumn}
\begin{document}

\thesaurus{ (08.09.2: 4U 1630-47; 13.25.5)}

\title{The 1998 outburst of the X-ray transient
\src\ observed with \sax}

\author{T.~Oosterbroek\inst{1} 
\and A.N.~Parmar\inst{1}
\and E. Kuulkers\inst{2}
\and T. Belloni\inst{3} 
\and M. van der Klis\inst{3} 
\and F. Frontera\inst{4} 
%\and M.~McCollough\inst{5,6} % ????
\and A.~Santangelo\inst{5}}

\institute
{Astrophysics Division, Space Science Department of ESA, 
ESTEC, P.O. Box 299, NL-2200 AG Noordwijk, The Netherlands
\and Astrophysics, University of Oxford, Nuclear and Astrophysics Laboratory,
Keble Road, Oxford, OX1 3RH, UK
\and Astronomical Institute, University of Amsterdam \& Center for
High-Energy Astrophysics, Kruislaan
403, NL-1098 SJ Amsterdam, The Netherlands
\and Istituto TESRE, CNR, via Gobetti 101, I-40129 Bologna, Italy
%\and ES-84, Space Science Laboratory, NASA/Marshall Space Flight
%Center, Huntsville, Alabama 35812, USA
%\and Universities Space Research Association
\and IFCAI, CNR, via La Malfa 153, I-90146 Palermo, Italy
}
\offprints{T. Oosterbroek: toosterb@astro.estec.esa.nl}
\date{Received 10 August 1998 / Accepted 30 September 1998}
%\titlerunning{T. Oosterbroek  et al., \src}{T. Oosterbroek  et al., \src}
%\authorrunning{T. Oosterbroek et al.}{T. Oosterbroek et al.}
\maketitle

\markboth{T. Oosterbroek et al.: \src}{T. Oosterbroek et al.: \src}

\begin{abstract}
We report results of five pointed \sax\ observations of the
black hole candidate \src\ during part of its 1998 outburst when
the 2--10~keV source luminosity decreased from
$2.1 \times 10^{38}$~\ergsec\ (for an assumed distance
of 10~kpc) by a factor of $\sim$7.
The 2--200~keV spectrum can be modeled by
a soft multi-temperature disk blackbody and a hard
power-law. During four of the observations, there is evidence for 
deviations from a pure power-law shape $\approxgt$20~keV which
may be due to reflection.
As the outburst decayed a number of spectral trends are
evident: (1) the amount
of photoelectric absorption decreased, (2) 
the spectrum became harder (i.e. the power-law index became smaller),
%(3) the relative contribution of the reflected
%component increased, 
(3) the temperature of the disk blackbody decreased
and (4) the inferred inner disk radius increased by a factor of $\sim$4. 
The change in the accretion disk inner radius during the outburst
is in contrast to results from other soft X-ray transients.
We find that the disk blackbody temperature depends on
the inner disk radius roughly as ${\rm T\propto r^{-3/4}}$. 
For a standard Shakura-Sunyaev disk model this implies that the 
inferred mass accretion rate at the inner disk radius 
did not change appreciably during the part of the outburst
observed by \sax.

\end{abstract}


\section{Introduction}
\label{sec:introduction}
\src\ is an X-ray transient discovered by {\it Uhuru} (Jones et al.\
1976), while the first recorded outburst was detected in 
1969 by {\it Vela} 5B (Priedhorsky 1986).
Outbursts occur approximately periodically with intervals
ranging from 600--690 days (Jones et al.\ 1976; Priedhorsky 1986; 
Kuulkers et al.\ 1997a). 
Archival observations are 
reported in Parmar et al. (1995, 1997b). 
The 1984 outburst, and its subsequent decay, were relatively well
studied by EXOSAT (Parmar et al. 1986). The flux decayed over a
period of $\sim$100 days from $\sim$2.8$\times10^{38}$ to
4$\times10^{36}$ ergs s$^{-1}$ (1--50 keV, for an assumed distance of
10~kpc). This distance estimate is based
on the high absorption towards the source and its proximity to the
galactic center (Parmar et al. 1986). 
The relative strength of the soft component decreased by at
least a factor of two during the decay and the power-law component
became harder as the outburst progressed.

\src\ is a black hole candidate based on its
X-ray spectral and timing behavior (White et al.\ 1984;
Parmar et al.\ 1986; Kuulkers et al. 1997b). 
The typical recurrence timescale for black hole candidate 
outbursts is 10--50~yr (e.g., Parmar et al. 1995), 
and the more prolific outburst activity 
of \src\ is therefore unusual.
No optical counterpart is known, probably due to high reddening and the
crowded field (see Parmar et al. 1986).
Kuulkers et al. (1997a) predicted that \src\ would 
undergo another outburst
near 1998 January 31. It went into outburst on around 1998 February 2,
and was detected by the All Sky Monitor (ASM) on-board RXTE 
(Kuulkers et al.\ 1998a). The outburst reached
a peak 2--12~keV intensity of $\sim$460 mCrab on around
February 24. During the outburst
a radio counterpart was detected by Hjellming \& Kuulkers (1998).

We report on five \sax\ pointed observations performed
during the decay of the 1998 outburst. We discuss the 2--200~keV
spectrum of \src\ and its evolution as the luminosity changed by a 
factor $\sim$7. In addition, we report on an observation in 
1997 March when the source was in quiescence.

\section{Observations and Analysis}
\label{sec:observations}
\begin{figure}
%\centering{\epsfig{figure=/usr4/users/toosterb/1630-47/lc2.ps,width=8cm}}
%\caption[]{The light curve of \src\ obtained from the MECS
%data}
\centering{\epsfig{figure=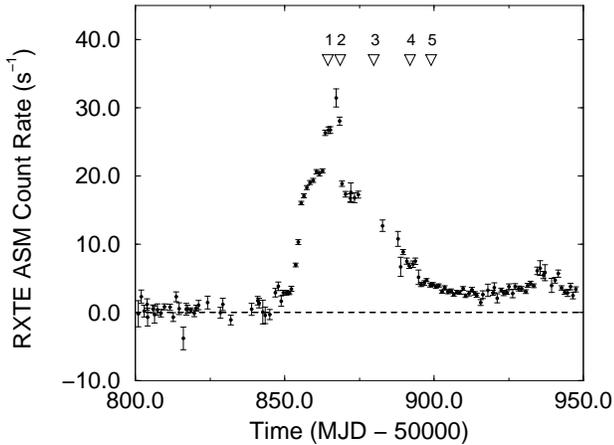,width=8cm}}
\caption[]{The RXTE ASM 2--12~keV light curve of the 1998 outburst
from \src. The times
of the \sax\ observations are indicated with open triangles. The
dashed line indicates the zero level}
\label{fig:lightcurve}
\end{figure}

\src\ was observed five times by \sax\ during the 1998 February-May 
outburst (Table \ref{tab:log}, see also Fig.\ \ref{fig:lightcurve}).
Results obtained 
with the coaligned LECS, MECS, HPGSPC,
and PDS instruments are presented. 
The LECS is sensitive between
0.1--10~keV (Parmar et al.\ 1997a), the MECS between 2--10~keV
(Boella et al.\ 1997), the HPGSPC between 4--120~keV
(Manzo et al. 1997), and the PDS between 15--300 keV
(Frontera et al.\ 1997). The LECS and the MECS are imaging
instruments with circular fields of view (FOVs) with diameters of
37\arcmin\ and 56\arcmin, respectively.
The non-imaging HPGSPC consists of a single unit with a collimator
that is alternatively rocked on- and off-source to monitor the background
spectrum.
The non-imaging PDS consists of four independent units arranged in
pairs each having a separate collimator.
Each collimator is operated in a rocking-mode to monitor the
background. The HPGSPC and PDS were operated in their nominal modes
with a dwell time of 96~s for each on- and off-source
position and rocking angles
of 210\arcmin\ and 180\arcmin, respectively.
The HPGSPC and PDS have hexagonal FOVs of
78\arcmin\ and 66\arcmin\ full-width at half maximum, respectively. 

\begin{table}
\caption[]{Log of the \sax\ observations of \src\ during its 1998 outburst}
\begin{tabular}{llllll}
\hline
Obs & Date & Start Time & Total & MECS  & MECS\\
    & \mc{1}{c}{} &  \mc{1}{c}{} & \mc{1}{l}{Time} &
    \mc{1}{l}{Exp.} & \mc{1}{l}{Count}\\
    & \mc{1}{c}{(1998)} &  \mc{1}{c}{(MJD)} & \mc{1}{l}{(ks)} &
    \mc{1}{l}{(ks)} & \mc{1}{l}{rate (s$^{-1}$)}\\
\hline
1 & Feb 20   & 50864.188 & 42.5 & 20.0 & 78.9 \\
2 & Feb 24   & 50868.254 & 34.1 & 13.5 & 75.7 \\
3 & Mar 07   & 50879.517 & 55.7 & 29.5 & 41.8 \\
4 & Mar 19   & 50891.620 & 48.4 & 27.7 & 20.8 \\
5 & Mar 26   & 50898.720 & 60.0 & 31.1 & 11.7 \\
\hline
\end{tabular}
\label{tab:log}
\end{table}

For the spectral fits only MECS data in the
energy range 1.85--10.5~keV and PDS data in the energy range
13--100~keV were used, since
-- due to the high absorption towards \src\ -- there is no useful
LECS data at lower energies, while the LECS spectrum in the 2--10~keV
band is consistent with that from the MECS. 
Typically, the source is detected with
more than 10$\sigma$ significance up to 200 keV in the PDS. 
The PDS spectra were
grouped using a logarithmic energy grid and the MECS spectra were
grouped such that each energy bin has at least 20 counts. The relative
normalization between the instruments was a free
parameter to allow for small absolute calibration uncertainties, 
but the best-fit values are always close to unity.

The HPGSPC data were used to check the spectral shape of the
source in the band 10.5--13 keV, which is not covered by the MECS and
PDS instruments. The shape obtained from the MECS and PDS
is in good agreement with the HPGSPC results.
The HPGSPC data were not used in the final fits, because there are systematic
residuals present which may be instrumental in
origin. These tend to increase the
values of \chisq\ (while the best-fit parameters remain unchanged).
These residuals may be partly caused by the non-optimal
background subtraction (see below) due to the presence of a source in
the offset collimator position.

The presence of contaminating sources in the off-source
collimator positions of the 
HPGSPC and the PDS was checked. 
During the later observations there is clearly
a source (probably GX~340+0) in the offset position of the
HPGSPC, and in the negative offset position of the PDS. Therefore,
data obtained during Earth occultations were used to accumulate 
background spectra
for the HPGSPC, and only the positive offset position was used 
for the PDS. Due to
the different roll-angle of the satellite for each observation,
the contribution of this contaminating source is
not significant during observations 1 and 2, but in order to obtain
a uniform dataset, the same method of background subtraction 
was used for 
all five observations.
Note that using only one offset spectrum in the background subtraction
results in a poorer signal to noise ratio, since the exposure time of the
background spectrum is effectively halved. 

\begin{table*}
\caption[]{The best-fit parameters for a power-law and disk
blackbody model. Uncertainties are given at 90\% confidence for one
parameter of interest. \nh\ is in units of $10^{22}$ atoms cm$^{-2}$}
\begin{tabular}{lllllll}
\hline
Obs & \nh & $\alpha$ & Norm$^{1}$ & T (keV) & Norm$^{2}$ & \rchisq/dof \\
    &     &          & (PL)     &         &  (Disk BB) \\
\hline
1 & 10.33$\pm0.08$ & 2.67$\pm0.01$ & 13.3$\pm0.5$ & 2.21$\pm0.03$ & 13.5$\pm1.3$ & 1.706/246  \\
2 & 9.88$\pm0.10$ & 2.70$\pm0.2$ & 11.3$\pm0.7$ & 1.96$\pm0.02$ & 27.2$\pm2.0$ & 0.993/246  \\
3 & 9.00$\pm0.06$ & 2.485$\pm0.01$ & 5.36$\pm0.02$ & 1.13$\pm0.03$ & 111$\pm12$ & 1.631/246  \\
4 & 8.32$\pm0.08$ & 2.30$\pm0.02$ & 1.10$\pm0.08$ & 1.020$\pm0.015$ & 218$\pm17$ & 1.177/246  \\
5 & 8.13$\pm0.10$ & 2.28$\pm0.03$ & 0.75$\pm0.05$ & 0.86$\pm0.02$ & 263$\pm30$ & 1.561/246  \\
\hline
\mc{7}{l}{\footnotesize $^{1}$Photon flux at 1 keV}\\
\mc{7}{l}{\footnotesize $^{2}$${\rm(R_{in}/D_{10})^2
\cos{\theta}}$, where ${\rm R_{in}}$ is the inner disk radius, 
${\rm D_{10}}$ the distance in units of}\\
\mc{7}{l}{\footnotesize 10~kpc and ${\rm \theta}$ the disk inclination
angle}\\
\end{tabular}
\label{tab:fits_pow}
\end{table*}

\begin{table*}
\caption[]{The best-fit parameters for a {\sc pexriv} and disk
blackbody model. All uncertainties are obtained with
$\Delta\chi^{2}=2.706$ (90\% interval for one parameter of interest).
\nh\ is in units of $10^{22}$ atoms cm$^{-2}$. L is the 2--10~keV
unabsorbed luminosity in units of $10^{38}$~\ergsec\ for an assumed distance
of 10~kpc. ${\rm L_{BB}/L}$ is the ratio of luminosity in the disk
blackbody compared to the total in the 2--10~keV energy range}
\begin{tabular}{llllllllll}
\hline
Obs & \nh & $\alpha$ & Norm$^{1}$ & Relative 
& T (keV) & Norm$^{1}$ &  L &${\rm L_{BB}/L}$ & \rchisq/dof \\
    &     &          & ({\sc pexriv})  & reflection$^{2}$   
&         & (Disk BB)&    &                 & \\
\hline
\noalign {\smallskip}
1 & 10.03$^{+0.21}_{-0.10}$ & 2.56$^{+0.12}_{-0.04}$ & 10.9$^{+2.0}_{-0.7}$ 
& $<$0.19 & 2.10$^{+0.10}_{-0.03}$ & 17.9$^{+1.8}_{-3.1}$ &
2.08 & 0.29 & 1.414/244 \\
2 & 9.88$^{+0.05}_{-0.11}$ & 2.70$\pm0.01$ & 11.3$\pm0.4$ 
& $<$0.06& 1.96$^{+0.02}_{-0.01}$ & $27.2 \pm 2.2$ &
2.00 & 0.34 & 1.002/244 \\
3 & 9.03$^{+0.06}_{-0.11}$ & 2.50$^{+0.02}_{-0.05}$ & 5.51$^{+0.20}_{-0.55}$ & $0.07 \pm 0.05$ & $1.13 \pm 0.03$ & 111.0$^{+13.5}_{-9.5}$
& 1.05 & 0.22 & 1.608/244 \\
4 & 8.39$^{+0.05}_{-0.10}$ & 2.35$^{+0.03}_{-0.05}$ & 1.22$^{+0.07}_{-0.13}$ & 0.18$^{+0.08}_{-0.11}$ & 1.01$^{+0.02}_{-0.01}$ & $225 \pm 17$
& 0.47 & 0.56 & 1.151/244 \\
5 & 8.23$\pm0.16$ & 2.37$^{+0.07}_{-0.12}$ & $0.79 \pm 0.15$ & 0.69$^{+0.23}_{-0.19}$ &0.85$\pm0.02$ &
282$^{+40}_{-31}$ & 0.27& 0.47 & 1.112/244\\
\noalign {\smallskip}
\hline
\mc{9}{l}{\footnotesize $^{1}$Units are the same as in 
Table \ref{tab:fits_pow}}\\
\mc{9}{l}{\footnotesize $^{2}$Constrained to be positive}\\
\end{tabular}
\label{tab:fits_pexriv}
\end{table*}

\section {Results}
\label{sec:analysis}
%\subsection{spectral analysis}

\begin{figure*}
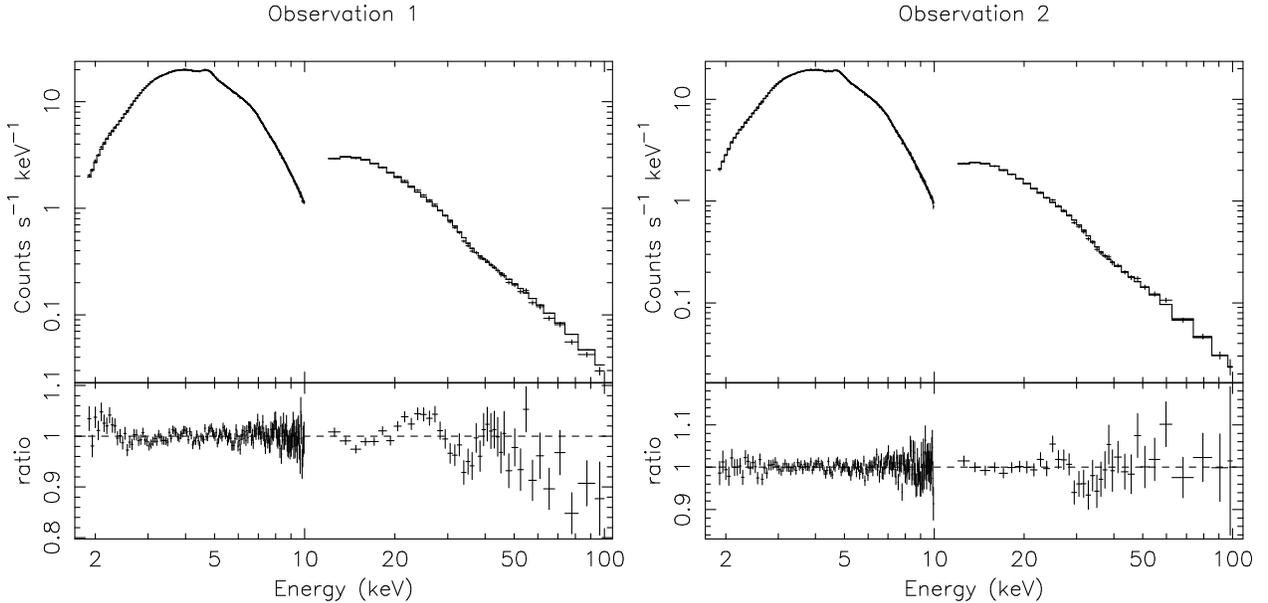

%\centering{\mbox{\epsfig{figure=/usr4/users/toosterb/1630-47/spec_1.ps,width=6cm,angle=-90}
%\epsfig{figure=/usr4/users/toosterb/1630-47/spec_2.ps,width=6cm,angle=-90}}}
\centering{\mbox{\epsfig{figure=obs1.ps,width=8cm,angle=-90}
\epsfig{figure=obs2.ps,width=8cm,angle=-90}}}
\caption[]{A comparison of MECS and PDS spectra with strong 
(left, obs.\ 1) and weak (right, obs.\ 2)
``wiggles'' at higher energies. 
The fit model is an absorbed power-law and disk blackbody}
\label{fig:wiggles}
\end{figure*}

A model consisting of an absorbed  multi-temperature disk
blackbody (Mitsuda et al.\ 1984) and an
absorbed power-law fits the general broad-band spectral 
shape reasonably
well (Table~\ref{tab:fits_pow}). However, with the exception of the
spectra obtained during observation 2, deviations from the best-fit model 
are present which
give rise to formally unacceptable values of \rchisq. Fitting the
spectra with a combination of absorbed power-law and blackbody models
gives similar results. The spectra were next fitted
with an absorbed disk blackbody and a cut-off
power-law model. This gives some improvement in fit quality,
but is still unable to satisfactorily model the 
deviations, or ``wiggles'', in the spectrum which are most prominent
around $\sim$20~keV (see Fig.\ \ref{fig:wiggles}). 
 
\begin{figure}
\centering{\epsfig{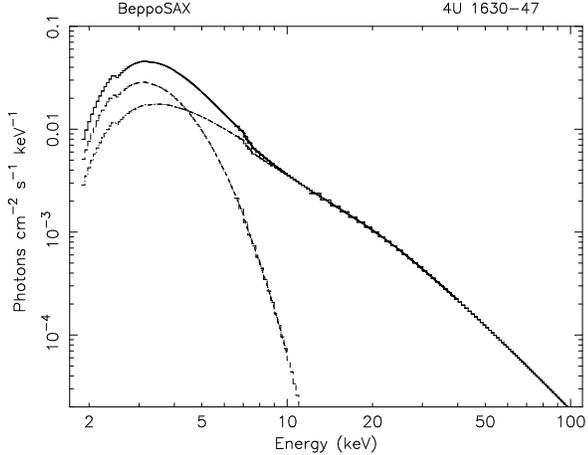}}
\caption[]{The photon spectrum of \src\ obtained during observation
5 with the MECS, HPGSPC, and PDS instruments. The disk blackbody
and power-law-like {\sc pexriv} components are shown seperately}
\label{fig:spectrum1}
\end{figure}

These deviations prompted the use of a model which
incorporates reflection such as {\sc pexriv} 
in XSPEC which models a power-law reflected by
ionized material (Magdziarz \& Zdziarski 1995).
The spectra were fit with the photon index, folding energy, 
relative reflection, and normalization as free parameters. 
The best-fit parameters are
given in Table~\ref{tab:fits_pexriv}, with the exception of the
folding energy, which was always much larger than highest energy used in
the fit. The ionization parameter, $\xi$, was fixed at 1.0
(corresponding to un-ionized material), since 
the fits are 
insensitive to changes in this parameter.
The effects of reflection can be seen as
deviations from a power-law continuum starting at $\sim$10
keV and extending to 30--40~keV in Fig.~\ref{fig:spectrum1}.
With this model the overall \rchisq\ is significantly less
than with a power-law and disk blackbody model, 
but individual values are still are not always formally acceptable. 
In comparison with
the cutoff power-law model, the values of \rchisq\
are similar for observation 1, while for
observations 2--4 the {\sc pexriv} model gives marginally 
lower values of \rchisq\ (at 
90--94\% confidence), and for observation 5 the reduction in
\rchisq\ is highly significant ($>$99.9999\% confidence). 
The trends
in \nh, disk blackbody temperature, and inner
disk radius are very similar whichever model is used for the hard
component (see e.g. Tables \ref{tab:fits_pow} and
\ref{tab:fits_pexriv}). Therefore the conclusions 
in Sect.~\ref{sec:discussion} with regard to the
derived disk blackbody parameters do not depend on the assumed hard-component
model.

%\begin{figure}
%\centering{\epsfig{figure=/usr4/users/toosterb/1630-47/spectrum.ps,width=6.5cm,angle=-90}}
%\caption[]{The photon spectrum of \src\ obtained during observation
%2. The disk blackbody and the power-law-like {\sc pexriv}) are visible. Note
%the large effects of the absorption by neutral material}
%\label{fig:spectrum1}
%\end{figure}

Tables~\ref{tab:fits_pow} and \ref{tab:fits_pexriv} reveal that
the temperature of the disk blackbody decreased
and the inferred inner disk radius increased as the outburst
progressed (see also Fig.~\ref{fig:param}).  
To investigate whether these changes could be
artefacts of the fitting procedure, the correlations
between the disk blackbody normalization (proportional to the square of the
inner disk radius) and the power-law slope and disk blackbody
temperature were investigated. 
Large changes are apparent in these quantities
and a (very) strong correlation between parameters could give rise
to an apparent change in the inner disk radius (which should then not
be interpreted in a physical way). 
In order to check that this is not the case,  
contour-plots corresponding to large changes in \chisq\
($\Delta$ \chisq=10, 20) were produced.
The confidence contours do
not overlap for the different observations since the differences between the
obtained parameters are an order of magnitude larger than the size
of the confidence contours for these already large \chisq differences.
This analysis reveals that 
the observed changes are {\it much} larger than can be
explained as due to correlations between fit parameters. The observed
changes in the spectral parameters therefore reflect physical changes.

Power spectra of \src\ were searched for any distinctive features such
as quasi-periodic oscillations. None were found. However,
the quality of the power spectra is rather poor (a much better
analysis can be done with e.g.\ RXTE data) and so a more
detailed timing analysis was not attempted.
%\begin{figure}
%\centering{\epsfig{file=/usr4/users/toosterb/1630-47/contour1.ps,width=6cm,angle=-90}}
%\vspace{0.3cm}\\
%\centering{\epsfig{file=/usr4/users/toosterb/1630-47/contour2.ps,width=6cm,angle=-90}}
%\caption[]{The error contours for the disk blackbody normalization and
%power-law index (top) and disk blackbody normalization and the
%temperature at the inner disk (bottom). 
%The small cross in both plots indicate the best-fit position (??!!).
%Both plots have been made with
%the data obtained from observation 2. {\bf I don't think we need those
%plots}}
%\label{fig:contours}
%\end{figure}

During the \sax\ observation of \src\ in its quiescent state  (on 1997
March 26--27), \src\ was not detected. However, due to the presence of
a number of nearby sources  and the relatively poor LECS and MECS
spatial resolution, it is difficult to separate the contributions
from nearby faint sources.
The 3$\sigma$ upper limit to the MECS count rate is
1.2$\times10^{-3}$~s$^{-1}$. This translates into an upper limit on
the 2--10 keV unabsorbed flux of  1.9$\times10^{33}$ ergs s$^{-1}$,
for an assumed distance of 10 kpc, a photon index of 2.1, and an
\nh\ of 8$\times10^{22}$ \hcm\ (close to lowest value obtained). This
photon index is the value obtained for the black
hole candidate GS\thinspace2023+338 in quiescence by Narayan et
al.\ (1997). The corresponding value for the unabsorbed 2--10~keV flux
for an \nh\ of 2.0$\times10^{22}$ \hcm (the galactic value; Dickey \&
Lockman 1990) is 1.0$\times10^{33}$ ergs s$^{-1}$.  This upper limit is
similar to previous limits for the flux of \src\ in quiescence
(see Parmar et al.\ 1997b).

\section {Discussion}
\label{sec:discussion}

\begin{figure}
\centering{\epsfig{figure=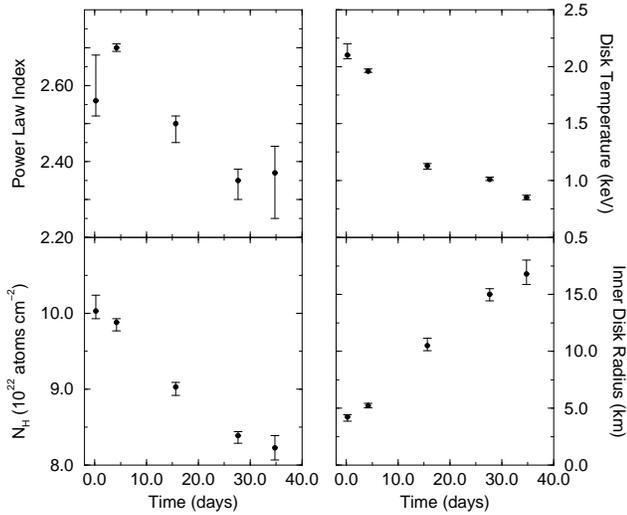,width=8.5cm}}
\caption[]{The changes as a function of time (in days)  in the best
fit parameters (see Table \ref{tab:fits_pexriv}). The uncertainties
correspond to 90\% confidence intervals. The inner disk radii
were obtained assuming a distance of 10~kpc and ${\rm \theta = 0\degmark}$}
\label{fig:param}
\end{figure}

The \sax\ observations of \src\ during its 1998 outburst reveal a
decreasing \nh, a hardening of a power-law like component, a
decreasing disk blackbody temperature, and an increasing 
inferred inner disk radius (Fig.~\ref{fig:param}).
These first two changes were also observed during the 1986
outburst of \src\ by Parmar et al. (1986), and in other
X-ray transients (e.g.\ GS\thinspace2023+338; Oosterbroek et al.\ 1997). The
change in \nh\ may suggest that the initial outburst onset is a rather 
dramatic event, which leaves significant amounts of material around the
compact object, which later on is either accreted, or expelled from the
system.

\begin{figure}
\centering{\epsfig{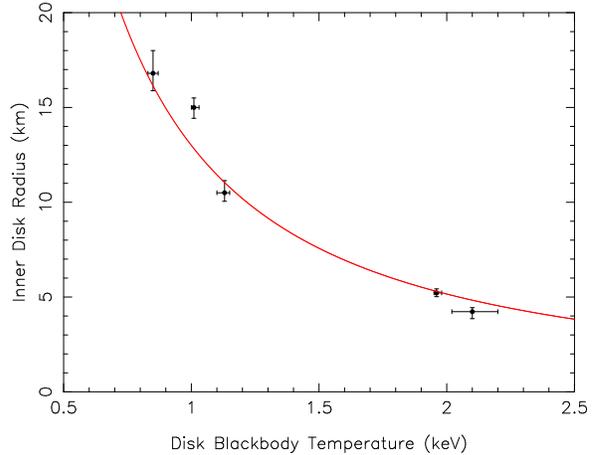}}
\caption[]{The relation between the inner disk temperature and the
best-fit inner disk radius. 
The inner disk radii
were obtained assuming a distance of 10~kpc and ${\rm \theta = 0\degmark}$.
Points are obtained (in time) from right
to left.
The solid line represents the theoretical
prediction 
for a standard Shakura-Sunyaev disk model (which is only
valid for $r>>R_{\rm in}$)}
\label{fig:temp_radius}
\end{figure}

The inner disk radius (which has been interpreted as the
radius of the innermost stable orbit around a black hole; Ebisawa
1991) increased by a factor of 4 (Fig.\ \ref{fig:param}) during
the part of the outburst observed by \sax.  Therefore it is clearly
not related (at least in a simple way) to the  innermost stable
orbit. It is interesting to note that a similar phenomenon, as seen in
GRS\thinspace1915+105 by Belloni et al.\ (1997), may be evident here:
the inner part of the accretion disk changes its state such that it
does not contribute to the X-ray flux. However, in
GRS\thinspace1915+105 this occurs on a much shorter time scale of
seconds to minutes. The fact that the best-fit temperature at the inner
disk radius decreased during the outburst supports this
interpretation - it is just the result of the change in effective
inner disk radius.  The relation between the best-fit inner disk radii
and temperature is shown in Fig.~\ref{fig:temp_radius}.
Comparing this to the theoretical temperature
profile of a standard Shakura-Sunyaev disk (${\rm T_{s}\propto
r^{-3/4}}$, where ${\rm T_{s}}$ is the surface temperature and r the
radius), shows a qualitive agreement. We note that this relation
is, strictly speaking, only valid  for radii much larger
than the inner disk radius.  Since the mass
accretion rate, ${\rm \dot{M} = 8\pi R_{\rm in}^{3}\sigma
T^{4}/3GM}$ this implies that ${\rm \dot{M}}$ is approximately constant
during the part of the outburst observed by \sax\ 
(if the contribution from the 
non-thermal component is excluded), and that the inner part of
the accretion disk progressively changed its state to produce
the observed spectral changes.

This behavior is quite different from outbursts of the transients
GS\thinspace1124$-$68, GRO{\thinspace}J1655$-$40 and 
GS\thinspace2000+25 and from the persistent
source LMC{\thinspace}X-3 observed by {\it Ginga} and RXTE. In those
sources the  best-fit inferred inner disk radii remain approximately
constant during the entire outburst (GS\thinspace2000+25), or
until the late stages ($\approxgt$150~days) of the outburst
(see Tanaka \& Lewin 1995; Ebisawa et al.\ 1994; Sobczak et al.\ 1998). 
This difference  
is unlikely to be caused by the broad energy coverage of
\sax,  which allows the underlying power-law(-like) spectrum to be
well constrained, since the {\it Ginga} and RXTE energy ranges
extend far beyond where the disk blackbody contributes to the
flux. 
Another difference with respect to the above sources is
that their disk blackbody temperatures are $\approxlt$1~keV, 
compared to $\sim$2~keV with \src\ near the peak of
the outburst. A similar high temperature was observed in
GRS\thinspace1915+105 (Belloni et al.\ 1997).
We speculate that
the difference in spectral properties may be related to the
difference in outburst recurrence intervals between \src\ (1.6~yrs)
and the other soft X-ray transients which have typical recurrence intervals
$\approxgt$25~yrs. This may suggest that a different mechanism is
responsible for the outbursts in \src, which also gives rise to a different
behavior during the outburst. The fact that the inner part of the
accretion disk disappears slowly (and smoothly) suggests that the
matter accretes into the black hole (as opposed to the sudden phase
transition observed in GRS\thinspace1915+105). This, in turn, may
cause a drop in $\dot{M}$ at the inner edge, which is needed to cause
the end of  an outburst in irradiated disks (see King \& Ritter 1998).

The increase in the best fit inner disk radius is compatibible
with the model of Esin et al. (1997) who assume that accretion
takes place through a Shakura-Sunyaev disk 
beyond some truncation radius, but within this radius the
flow jumps to the advective solution. They model the change from a high
to a low state as an increase in the radius where the character
of the flow changes.  The observed increase in best-fit inner disk radius
may therefore simply reflect the outward expansion of the radius where 
optically thick material is present as the outburst evolved. 
We note that the best-fit values for the inner
disk radius are  lower than the value (3 Schwarzschild radii, the
innermost marginally stable orbit) for the high state in the 
Esin et al. (1997) model. However, 
we have assumed that the disk is viewed face-on (an inclination angle of
0$\degmark$), while there are indications
that the inclination is in the range 60--75$\degmark$ (Kuulkers et
al.\ 1998b). In addition, the distance to the source is uncertain. These
factors may
result in larger values for the ``real'' inner disk radius.  \.{Z}ycki
et al. (1998) also find much smaller values for the inner disk radius
from their modeling of the spectra of GS\thinspace1124$-$68, compared
to the values predicted in Esin et al.\ (1997).

We note that the changing inferred inner disk radius has
implications for the method applied by Zhang et al.\ (1997) to
determine black hole spin. They calculate the general relativistic
effects of black hole rotation on the inner disk regions, which allows
the obtained disk blackbody temperatures and inner disk radii to be
interpreted in terms of the spin of the black hole.  Our results
suggest that other processes also affect the spectrum (since a change
of black hole spin can be excluded on a time scale of $\sim$40 days),
which cautions against interpreting spectral properties in term of
black hole parameters alone.

The nature of the hard component in \src\ is unclear. Although
a reflection model provides the lowest overall \rchisq,
this does not necessarily mean that significant reflection is 
occuring in the system.
The largest reduction in \rchisq\ using the {\sc pexriv} 
model compared to a power-law is obtained when the fitted
relative reflection is high, which {\it does} mean that the deviations from a
power-law, to first order, can be modeled with a reflection
component. However, the values of \rchisq\ are not always formally acceptable,
which suggests that, although the {\sc pexriv} model gives a 
better overall fit then a power-law or a cutoff power-law, 
the detailed modeling of the
reflected component is unsatisfactory. 
We note that the fits to the EXOSAT 
spectra also gave formally unacceptable \chisq\ values, although
Parmar et al. (1986) used a different model consisting of a
Wien-like soft component and a power-law.  It appears that the
spectrum of \src\ is complex and contains features which cannot be
currently modeled in a realistic way.

The amount of reflection in the {\sc pexriv} model (see Table
\ref{tab:fits_pexriv}) increased, compared to the underlying continuum,
as the outburst decayed. 
However, there is no obvious
correlation between the amount of relative reflection and the
normalizations of the disk blackbody and
power-law, as modeled with the {\sc pexriv} model, suggesting that any
reflected component is neither constant nor directly related to the
strengths of either spectral component.
The weakness of the ``wiggles'' during observation 2 
coincided with the peak of the outburst, as can be seen from
the RXTE ASM light curve (Fig.\ \ref{fig:lightcurve}). This may be
coincidence, or it could be that the continuum becomes
``smoother'' near the maximum of the outburst as a rule. We note
that the residuals in the $\sim$4--7 keV range become significantly
less if a narrow Gaussian component is added, suggesting the presence of a
weak ($\sim$30~eV equivalent width) iron line (obs.\ 3). 
However, the residuals are at
the 2--3\% level, comparable to the systematic uncertainties.
In comparison with Active Galactic Nuclei where reflection components
and associated iron fluorescence lines have been detected from a 
number of systems (e.g., Pounds et al. 1990), the lack of
intense ($\approxgt$100~eV equivalent width) iron emission from \src\ 
is surprizing, if reflection is important in this system. 
However, we note that during a Broad Band X-ray Telescope observation
of the blackhole candidate Cyg{\thinspace}X-1 by Marshall et al. (1993)
also showed evidence for reflection by material surrounding the central
source, but only a narrow line at 6.4~keV with an equivalent width of
$13 \pm 11$~eV was possibly detected.

\begin{acknowledgements}
  The \sax\ satellite is a joint Italian--Dutch programme.
  T. Oosterbroek acknowledges an ESA Fellowship. We thank the staff
  of the \sax\ Science Data Center for their support and 
  the RXTE/ASM teams at MIT and at NASA's GSFC 
  for providing quick-look results.
\end{acknowledgements}

\end{document}